\documentclass[12pt]{article}

\usepackage[totalheight = 23cm, totalwidth = 17cm]{geometry}
\usepackage{amssymb,amsmath,amsfonts,amsbsy,epsfig,psfrag,rotating,chngpage}

\begin{document}

\title{End of multi-field inflation and the perturbation spectrum}

\author{
Jinn-Ouk Gong\footnote{jgong@kasi.re.kr}
\\ \\
\textit{International Center for Astrophysics}
\\
\textit{Korea Astronomy and Space Science Institute}
\\
\textit{Daejeon, Republic of Korea} }

\date{\today}

\maketitle

\begin{abstract}

We investigate the dynamics of inflation models driven by multiple, decoupled scalar
fields and calculate the Hubble parameter and the amplitude of the lightest field at
the end of inflation which may be responsible for interesting, or possibly dangerous
cosmological consequences after inflation. The results are very simple and similar
to those of the single field inflation, mainly depending on the underlying spectrum
of the masses. The mass distribution is heavily constrained by the power spectrum of
density perturbations $\mathcal{P}$ and the spectral index $n_\mathrm{s}$. The
overall mass scale gives the amplitude of $\mathcal{P}$, and $n_\mathrm{s}$ is
affected by the number of fields and the spacing between masses in the distribution.
The drop-out effect of the massive fields makes the perturbation spectrum typically
redder than the single field inflation spectrum. We illustrate this using two
different mass distributions.

\end{abstract}

\thispagestyle{empty}
\setcounter{page}{0}
\newpage
\setcounter{page}{1}

\section{Introduction}

After its advent, inflation \cite{infbook} has been very successful to solve many
cosmological problems of the standard hot big bang cosmology and its predictions are
consistent with the recent observations \cite{sdss,wmap3}: inflation can explain the
homogeneity, isotropy and flatness of the observable universe, as well as the origin
of the large scale structure in the universe. Interestingly, the simplest
possibility, i.e. the model of chaotic inflation with quadratic potential $V(\phi) =
m^2\phi^2/2$ is still consistent with the WMAP three year results \cite{wmap3}. Thus
it is natural to try to build such a fully realistic inflationary scenario in the
context of string theory, the most promising candidate of the fundamental theory of
nature. However, it is still unclear how to find successful inflationary scenarios
in string theory. Especially, if the vacuum expectation value of the inflaton field
$\phi$ is beyond the Planck scale $m_\mathrm{Pl}$, which is the case for the
simplest $m^2\phi^2$ chaotic inflation, generally large radiative corrections are
expected and they would modify the inflaton potential $V(\phi)$, and the predictions
based on the original form of the effective potential will be no more reliable.

Fortunately, after the de Sitter vacua construction in type IIB theory \cite{kklt},
there have been considerable advances to this end for recent years \cite{advances}.
An interesting scenario named N-flation was recently proposed \cite{dkmw} where the
$m^2\phi^2$ inflation model is achieved by the assisted inflation mechanism
\cite{assisted}: many ($N \gg 1$) fields combine to drive inflation with the
displacements being less than $m_\mathrm{Pl}$. These fields are considered to be the
string axion fields, which are known to exist abundantly in string theory, and they
have flat potentials even after all the moduli are stabilised. In Ref.~\cite{dkmw}
the special case where all the masses are the same is highlighted, and subsequently
more general extensions were investigated in Refs.~\cite{mpdist,kimliddle}. The
general and qualitative evolution of the fields is however clear: the most massive
fields roll down their potentials to the minima, and the light fields are
responsible for the later inflationary stages. Here one important thing is that, at
the end of inflation, {\em not} all the fields are relaxed to the minima. For the
single field inflation with $V(\phi) = m^2\phi^2/2$, as we will see in the next
section, inflation ends well before $\phi$ reaches the minimum of the potential.
Likewise, inflation proceeds no more although there are still some fields whose
amplitudes are non-zero. These non-zero fields are of cosmological importance after
inflation: they may lead to interesting, or possibly dangerous cosmological
consequences. For example, they may oscillate and decay into dark matter particles
\cite{dm}, they can produce additional perturbations via curvaton
mechanism\footnote{Note that for the curvaton mechanism to work properly, we must
put more constraints on the parameter space of the model under consideration
\cite{curvatonwork} in addition to the conditions for successful inflationary
phase.} \cite{curvaton}, and they would be responsible for the preheating
\cite{preheating} of the universe after inflation. They can, of course, produce
gravitino at an unacceptable level \cite{gravitino}, spoiling the successful
predictions of the big bang nucleosynthesis \cite{nucleo}. The dynamics of the
fields depends on the underlying spectrum of masses, which is revealed via the power
spectrum $\mathcal{P}$ and the spectral index $n_\mathrm{s}$: they depend on the
detail of the mass distribution.

Our purpose in this paper is to study the dynamics of this class of multiple field
inflation to calculate several interesting physical quantities at the end of
inflationary regime in connection with the power spectrum $\mathcal{P}$ and the
spectral index $n_\mathrm{s}$. Since the lightest field keeps the largest amplitude
at the end of inflation, we will mainly focus on this field and its mass: also we
can describe the evolution of the lightest field under slow-roll approximation, and
this helps to simplify the calculations. This paper is outlined as follows. In
Section~\ref{secsingle}, we first briefly review the simplest chaotic inflation
model with $V(\phi) = m^2\phi^2/2$ and calculate some quantities of the model. In
Section~\ref{secNflation}, we calculate the Hubble parameter, the number of fields
with non-zero amplitudes and the amplitude of the lightest field at the end of
inflation. In Section~\ref{secmassdist} we apply the results of the previous section
to two interesting mass distributions, and compare with the single field inflation
case. In Section~\ref{secconclusion} we summarise and conclude.

\section{Single field inflation}
\label{secsingle}

\begin{figure}[h]
\psfrag{Pn}{$\phi$}%
\psfrag{H}{$H$}%
\psfrag{P}{$\mathcal{P}$}%
\psfrag{ns}{$n_\mathrm{s}$}%
\psfrag{N}{$\mathcal{N}$}%
\begin{center}
\epsfig{file=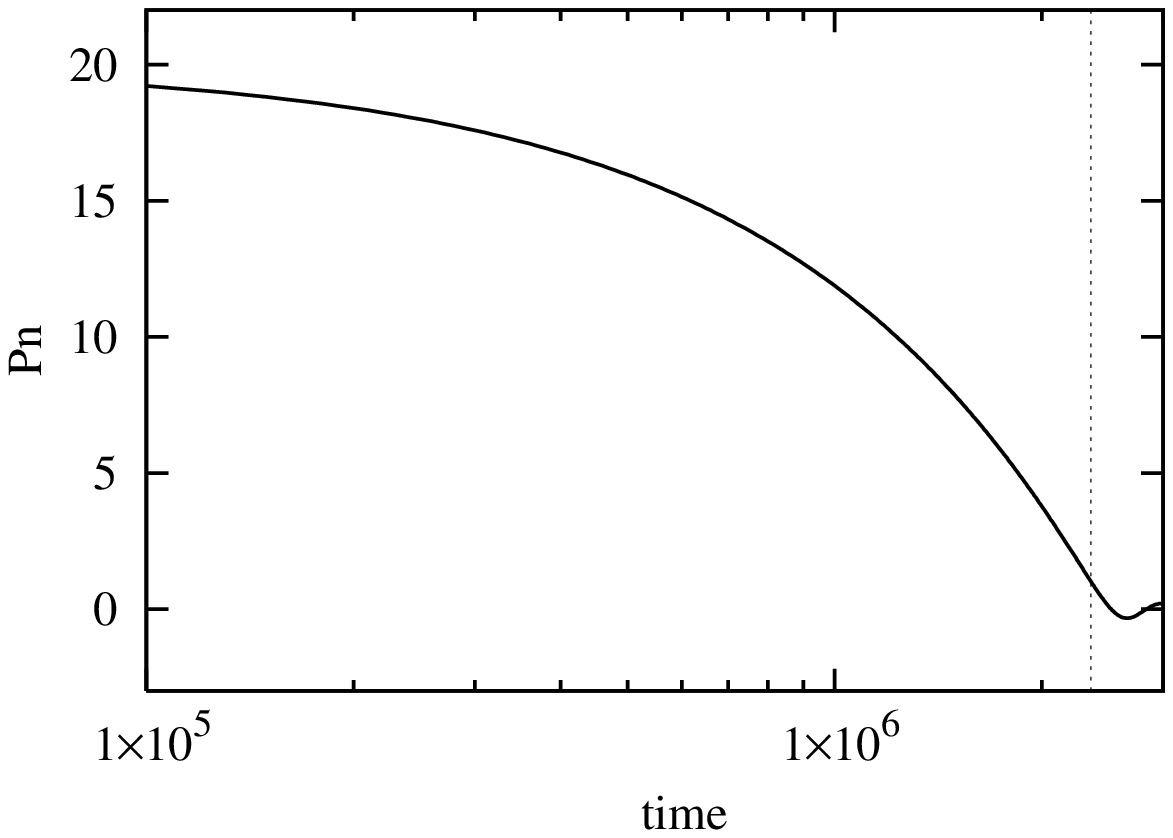, angle = -90, width = 8.1cm}%
\epsfig{file=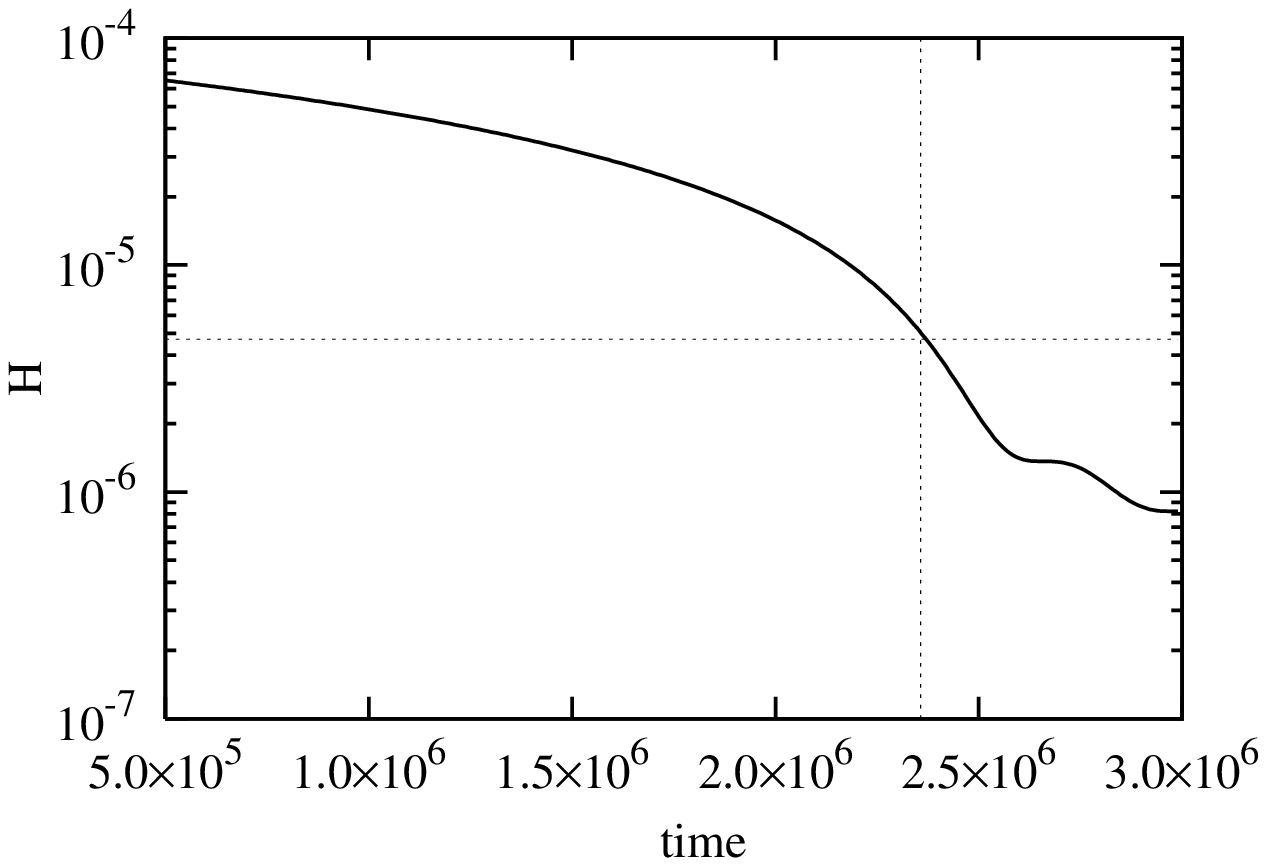, angle = -90, width = 8.1cm}%
\end{center}
\begin{center}
\epsfig{file=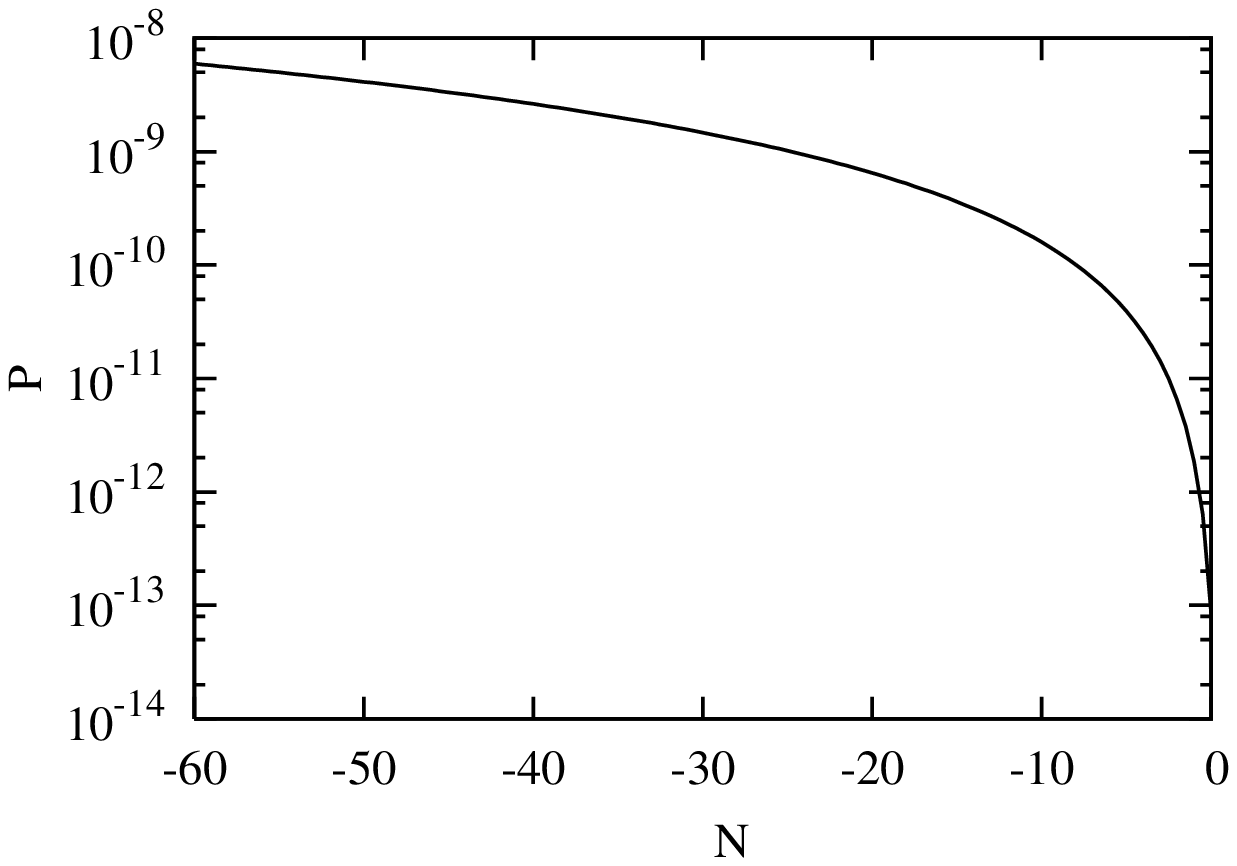, angle = -90, width = 8.1cm}%
\epsfig{file=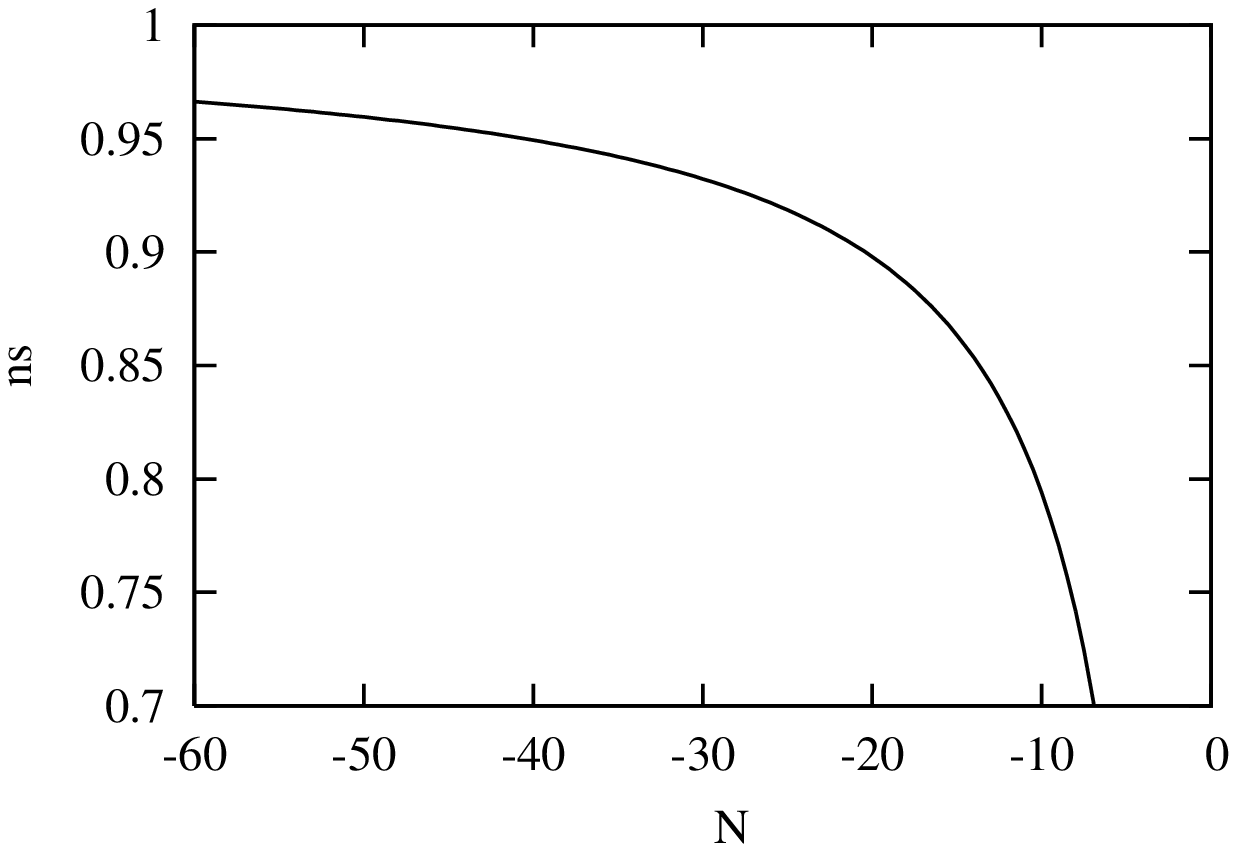, angle = -90, width = 8.1cm}%
\end{center}
\caption{Numerical results of the single field inflation with the initial amplitude
$\phi_i = 20 m_\mathrm{Pl}$ and the mass $m = 10^{-5} m_\mathrm{Pl}$. The dotted
lines indicate the end of inflation: in the plot of $H$ (upper right) the horizontal
dotted line is $H_\mathrm{end}$, given by Eq.~(\ref{singleHend}). In the lower row,
we set $\mathcal{N} = 0$ as the end of inflation.}%
\label{singleplot}
\end{figure}

In this section, we recall the inflation model where accelerated expansion is driven
by a single inflaton field with simple quadratic potential \cite{infbook}. The
inflaton potential is given by
\begin{equation}
V(\phi) = \frac{1}{2}m^2\phi^2 \, ,
\end{equation}
and with the slow-roll equation of motion for $\phi$,
\begin{equation}\label{singleSReq}
3H\dot\phi + m^2\phi = 0 \, ,
\end{equation}
the total number of $e$-folds is
\begin{equation}
\mathcal{N} = \int H dt \simeq \frac{\phi_i^2}{4m_\mathrm{Pl}^2} \, .
\end{equation}
Then, to obtain $\mathcal{N} \gtrsim 60$, we need $\phi_i \gtrsim
4\sqrt{15}m_\mathrm{Pl} \sim 15m_\mathrm{Pl}$: the initial value of the inflaton
$\phi$ is required to be much larger than $m_\mathrm{Pl}$. Now, from the
acceleration equation
\begin{align}\label{acceq}
\frac{\ddot{a}}{a} & = \frac{-1}{6m_\mathrm{Pl}^2} (\rho + 3p)
\nonumber \\
& = \frac{V(\phi) - \dot\phi^2}{3m_\mathrm{Pl}^2} \, ,
\end{align}
we find that inflation ends when
\begin{equation}\label{singlenoinf}
\dot\phi^2 \geq \frac{1}{2}m^2\phi^2 \, ,
\end{equation}
i.e. the kinetic energy of the field becomes greater than half of the potential
energy. Using Eq.~(\ref{singleSReq}) to eliminate $\dot\phi$, the value of the
Hubble parameter at the end of inflation is estimated as
\begin{equation}\label{singleHend}
H_\mathrm{end} = \frac{\sqrt{2}}{3}m \, .
\end{equation}
Note that Eq.~(\ref{singlenoinf}) is equivalent to the condition that the slow-roll
parameter
\begin{equation}\label{SRparameter}
\epsilon = -\frac{\dot{H}}{H^2} = \frac{1}{2m_\mathrm{Pl}^2} \left(
\frac{\dot\phi}{H} \right)^2
\end{equation}
becomes larger than 1. Substituting Eq.~(\ref{singleHend}) into
Eq.~(\ref{SRparameter}) and using Eq.~(\ref{singlenoinf}), we find that at the end
of inflation $\phi$ has the amplitude
\begin{equation}\label{singlephiend}
\phi_\mathrm{end} = \frac{2\sqrt{2}}{3}m_\mathrm{Pl} \, .
\end{equation}
The mass of the inflaton field is heavily constrained by the amplitude of the
primordial density perturbations\footnote{Note that the inflationary energy scale is
constrained from the detailed calculations on the amplitude of the power spectrum of
the primordial gravitational waves \cite{gw}.}. The leading-order\footnote{For the
calculations with higher-order corrections in the slow-roll parameters, see e.g.
\cite{higherorder}.} power spectrum of primordial density perturbations is given by
\cite{infbook}
\begin{equation}
\mathcal{P} = \frac{m^2\phi^4}{96\pi^2m_\mathrm{Pl}^6} \, ,
\end{equation}
and the corresponding spectral index by
\begin{equation}
n_\mathrm{s} - 1 = -\frac{8m_\mathrm{Pl}^2}{\phi^2} \, .
\end{equation}
Using the COBE observation $\delta_H = 2\sqrt{\mathcal{P}}/5 \simeq 1.91 \times
10^{-5}$, we find
\begin{equation}
m \simeq 6.07 \times 10^{-6} m_\mathrm{Pl} \, .
\end{equation}
Note that $n_\mathrm{s}$ is independent of $m$ and is given by $n_\mathrm{s} \simeq
0.967$ when $\phi = 4\sqrt{15}m_\mathrm{Pl}$, i.e. 60 $e$-folds before the end of
inflation\footnote{Note that, in fact, there exist some level of uncertainty on
exactly when the observable large scale perturbations were generated. While for most
models of inflation the corresponding $e$-fold lies between 50 and 60
\cite{uncertainty}, the values of $\mathcal{P}$ and especially $n_\mathrm{s}$ are
similar: we have $n_\mathrm{s} = 0.963280$ and 0.959581 for 55 and 50 $e$-folds
before the end of inflation, respectively. Thus, in this paper we keep evaluating
$\mathcal{P}$ and $n_\mathrm{s}$ before 60 $e$-folds.}. In Fig.~\ref{singleplot}, we
illustrate some results of numerical calculations, and in Table~\ref{comparetable}
we compare the corresponding analytic and numerical estimates.

\begin{table}[h]
\begin{adjustwidth}{-4em}{-4em}
\begin{center}
\begin{tabular}{r|r||*{4}{l|}l}
    \multicolumn{2}{r||}{} & $\mathcal{N}$ & $\phi_\mathrm{end}$ & $H_\mathrm{end}$
    & $\mathcal{P}$ & $n_\mathrm{s}$
    \\
    \hline\hline
    single field & analytic & 100.00 & 0.942809 & $0.471405 \times 10^{-5}$ &
    $6.18101 \times 10^{-9}$ & 0.966942
    \\
    \cline{2-7}
    inflation & numerical & 100.97 & 1.00934 & $0.574670 \times 10^{-5}$ & $5.96853
    \times 10^{-9}$ & 0.966359
    \\
    \hline
    exponential & analytic & 100.00 & 0.121716 & $0.101500 \times 10^{-5}$ &
    $2.47245 \times 10^{-9}$ & 0.936631
    \\
    \cline{2-5}
    distribution & numerical & 101.43 & 0.225956 & $0.124816 \times 10^{-5}$ & &
    \\
    \hline
    Mar{\v c}enko-Pastur & analytic & 99.99 & 0.175075 & $0.141009 \times 10^{-5}$ &
    $4.40243 \times 10^{-9}$ & 0.955755
    \\
    \cline{2-5}
    distribution & numerical & 101.27 & 0.324082 & $0.190495 \times 10^{-5}$ & &
\end{tabular}
\end{center}
\end{adjustwidth}
\caption{Analytic and numerical estimates of several quantities of the single and
multiple field inflation models under the mass distributions, Eqs.~(\ref{logdist})
and (\ref{MPdist}). The quantities for the single field inflation are from
Fig.~\ref{singleplot}. Here $\phi_\mathrm{end}$ and $H_\mathrm{end}$ are written in
the unit of $m_\mathrm{Pl}$, and $\mathcal{P}$ and $n_\mathrm{s}$ are calculated at
60 $e$-folds before the end of inflation. Note that since it is not easy to derive
analytic estimates for $\mathcal{P}$ and $n_\mathrm{s}$ for multi-field cases, we
only show their numerical results. Generally the analytic results agree with the
corresponding numerical ones.}%
\label{comparetable}
\end{table}

\section{Multiple field case}
\label{secNflation}

Now let us consider the inflation case with multiple, uncoupled fields \cite{dkmw}.
Here inflation is driven by a number of non-interacting string axion fields. If each
axion field is displaced not too far from the minimum, we can write the potential as
\begin{equation}\label{NflationV}
V = \sum_{i=1}^N \frac{1}{2} m_i^2 \phi_i^2 \, ,
\end{equation}
where we assume that there are $N$ axion fields. The total number of $e$-folds is
then given by
\begin{equation}
\mathcal{N} = \frac{\sum_i {\phi_i^{(0)}}^2}{4m_\mathrm{Pl}^2} \, ,
\end{equation}
where $\phi_i^{(0)}$ is the initial value of $\phi_i$. From Eq.~(\ref{acceq}), we
obtain
\begin{equation}
\frac{\ddot{a}}{a} = \frac{1}{3m_\mathrm{Pl}^2} \sum_{i=1}^N \left( \frac{1}{2}
m_i^2 \phi_i^2 - \dot\phi_i^2 \right) \, ,
\end{equation}
thus inflation ends when
\begin{equation}\label{infcondition}
\sum_{i = 1}^N \dot\phi_i^2 \geq \sum_{i = 1}^N \frac{1}{2} m_i^2 \phi_i^2 \, .
\end{equation}
Now, using the slow-roll equation of motion for $\phi_i$,
\begin{equation}
3H\dot\phi_i + m_i^2\phi_i = 0 \, ,
\end{equation}
we can find that under the slow-roll approximation different fields $\phi_i$ and
$\phi_j$ satisfy the relation \cite{ko}
\begin{equation}\label{fieldrelation}
\frac{\dot\phi_i}{\dot\phi_j} = \frac{m_i^2\phi_i}{m_j^2\phi_j} \, ,
\end{equation}
thus we obtain
\begin{equation}
\left( \frac{\phi_i}{\phi_i^{(0)}} \right)^{1/m_i^2} = \left(
\frac{\phi_j}{\phi_j^{(0)}} \right)^{1/m_j^2} \, .
\end{equation}
Therefore, we can write $\phi_i$ and $\dot\phi_i$ in terms of the lightest field
$\phi_N$ and $\dot\phi_N$ as
\begin{align}\label{phiNrelation}
\phi_i & = \left( \frac{\phi_N}{\phi_N^{(0)}} \right)^{m_i^2/m_N^2} \phi_i^{(0)} \,
,
\nonumber \\
\dot\phi_i & = \frac{m_i^2}{m_N^2} \left( \frac{\phi_N}{\phi_N^{(0)}}
\right)^{m_i^2/m_N^2 - 1} \left( \frac{\phi_i^{(0)}}{\phi_N^{(0)}} \right)
\dot\phi_N \, ,
\end{align}
respectively. Substituting Eq.~(\ref{phiNrelation}) into Eq.~(\ref{infcondition})
and using the slow-roll equation of motion for $\phi_N$, we find that inflation ends
when
\begin{equation}\label{Hend}
H_\mathrm{end}^2 = \frac{2}{9} \frac{\sum_{i = 1}^N m_i^4 \left(
\phi_N^\mathrm{end}/\phi_N^{(0)} \right)^{2m_i^2/m_N^2}}{\sum_{i = 1}^N m_i^2 \left(
\phi_N^\mathrm{end}/\phi_N^{(0)} \right)^{2m_i^2/m_N^2}} \, .
\end{equation}
Because $\phi_N^\mathrm{end} / \phi_N^{(0)} < 1$ at the end of inflation and $m_i >
m_N$ always, the last term of each sum gives the most significant contribution for
that sum. Therefore, the Hubble parameter at the end of inflation is
\begin{equation}\label{Hendsimple}
H_\mathrm{end} \simeq \frac{\sqrt{2}}{3}m_N \, ,
\end{equation}
where $m_N$ is the mass of the lightest field. Interestingly, this is exactly the
same form as the single field inflation case, Eq.~(\ref{singleHend}). Note that at
this point, combining Eq.~(\ref{Hendsimple}) and the slow-roll equation of $\phi_N$,
the kinetic energy of the field $\phi_N$ is half of the potential energy,
\begin{equation}\label{KVforN}
K_N = \frac{1}{2}V_N \, ,
\end{equation}
as the case of the single field inflation. Likewise for $\phi_i$ we can derive a
similar relation
\begin{equation}\label{KVfori}
K_i = \frac{m_i^2}{2m_N^2}V_i \, .
\end{equation}
Now, assuming that total $n$ light fields ($n < N$) contribute at the end of
inflation, we write
\begin{align}\label{nfieldcontribution}
H_\mathrm{end}^2 & \simeq \frac{2}{9}m_N^2
\nonumber \\
& = \frac{1}{3m_\mathrm{Pl}^2} \left[ (K_{N - n + 1} + V_{N - n + 1}) + \cdots +
(K_N + V_N) \right] \, .
\end{align}
As can be seen from Eq.~(\ref{KVfori}), as the mass becomes heavier, the kinetic
energy is increasing while the potential energy is decreasing. Thus, we can assume
that the energy contribution of each field roughly coincides. Then,
Eq.~(\ref{nfieldcontribution}) becomes
\begin{equation}
H_\mathrm{end}^2 \sim \frac{n}{3m_\mathrm{Pl}^2} (K_N + V_N) \, ,
\end{equation}
hence using Eq.~(\ref{KVforN}) the amplitude of the lightest field $\phi_N$ at the
end of inflation is given by
\begin{equation}\label{phiNend}
\phi_N^\mathrm{end} \sim \frac{2\sqrt{2}}{3} \frac{m_\mathrm{Pl}}{\sqrt{n}} \, ,
\end{equation}
which is again similar to the single field case, Eq.~(\ref{singlephiend}), divided
by some numerical factor $n$. Note that this number would be proportional to the
total number of fields $N$ unless the masses are anomalously distributed\footnote{If
the axion decay constant could be made far larger than $m_\mathrm{Pl}$, we can
obtain consistent inflationary scenario with very limited number of fields, e.g.
only two fields \cite{2axions}.}, hence $\phi_N^\mathrm{end}$ decreases as $N$
increases: with larger number of fields, the inflationary phase lasts longer until
$\phi_N$ reaches smaller value. To estimate the mass of the most massive field which
contributes at the end of inflation, first we observe that heavy fields drop out at
the early stages of inflation while light fields drive inflation at late stages.
Hence, it is reasonable to assume that the fields which are responsible for the
inflation of the last $e$-folds have masses within far less than an order of
magnitude, otherwise some of them would have already quitted the inflationary
regime. Thus, we can set the mass of the heaviest field to be
\begin{equation}\label{heaviest}
m_{N - n + 1} \sim \sqrt{2} m_N \, .
\end{equation}
This number $n$ is dependent on the specific mass distribution as we will see in the
next section. We cannot, however, follow the same steps to calculate the power
spectrum of curvature perturbations\footnote{Generally, it is expected that
orthogonal isocurvature perturbations are also generated. See, e.g. \cite{multipert}
for a review.} and the spectral index: following $\delta\mathcal{N}$ formalism
\cite{deltaN}, they are written as \cite{mpdist,kimliddle}
\begin{equation}\label{spectrum}
\mathcal{P} = \frac{\sum_i m_i^2\phi_i^2}{96\pi^2m_\mathrm{Pl}^6} \sum_j \phi_j^2
\end{equation}
and
\begin{equation}\label{index}
n_\mathrm{s} - 1 = -4m_\mathrm{Pl}^2 \left[ \frac{\sum_i m_i^4\phi_i^2}{\left(
\sum_j m_j^2\phi_j^2 \right)^2} + \frac{1}{\sum_k \phi_k^2} \right] \, ,
\end{equation}
respectively. From Eq.~(\ref{index}), we can see that the spectrum is {\em always}
redder than the spectrum of the single field inflation\footnote{Note that this is
true not only for the quadratic potential given by Eq.~(\ref{NflationV}), but for a
set of general power law potentials \cite{Nflationpowerlaw}.}. At the end of
inflation, the amplitude of $\phi_N$ has significantly decreased from its initial
value $\phi_N^{(0)}$, and thus we could ignore all the contributions except the one
by $\phi_N$ in Eq.~(\ref{Hend}). But before 60 $e$-folds where the COBE observation
is made, $\phi_N$ is almost frozen and hence the contributions other than $\phi_N$
should be taken into account.

\section{Mass distributions}
\label{secmassdist}

In this section we illustrate the results of the previous section using two
different mass distributions of the axion fields. Here we take $\phi_1^{(0)} =
\cdots = \phi_N^{(0)} = m_\mathrm{Pl}$ for numerical calculations: when building
inflationary models with masses being given by the underlying physics, the one with
the largest possible number of $e$-folds is favoured because it occupies the
greatest volume in the context of eternally inflating universe \cite{eternal} and is
preferred a posteriori. This is implemented by assigning the maximum initial
displacement to the fields with a given spectrum of masses.

\subsection{Exponential distribution}
\label{subseclogdist}

We first consider the exponential mass spectrum where the fields are distributed
uniform on logarithmic scale \cite{dkmw,kimliddle},
\begin{equation}\label{logdist}
m_i^2 = m_1^2 e^{-(i-1)/\sigma} \, ,
\end{equation}
where $i = 1, \cdots, N$ is the index of the fields, and $\sigma$ is the density of
fields per logarithmic interval. $\sigma$ is determined once we decide the heaviest
mass $m_1$, the lightest mass $m_N$ and the number of fields $N$.

\begin{figure}[h]
\psfrag{Pn}{$\phi$}%
\psfrag{225}{$\phi_{225}$}\psfrag{230}{$\phi_{230}$}\psfrag{235}{$\phi_{235}$}
\psfrag{240}{$\phi_{240}$}\psfrag{245}{$\phi_{245}$}\psfrag{250}{$\phi_{250}$}%
\begin{center}
\epsfig{file=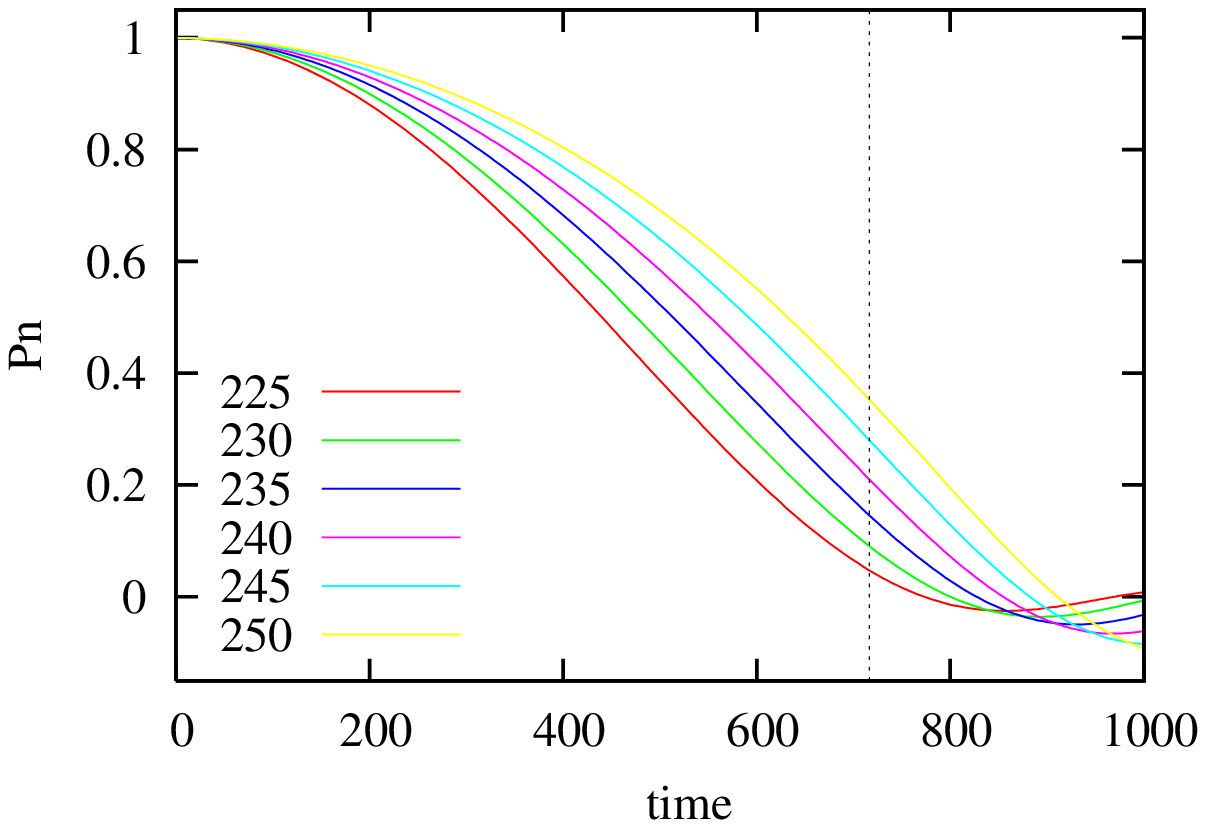, angle = -90, width = 8.1cm}%
\epsfig{file=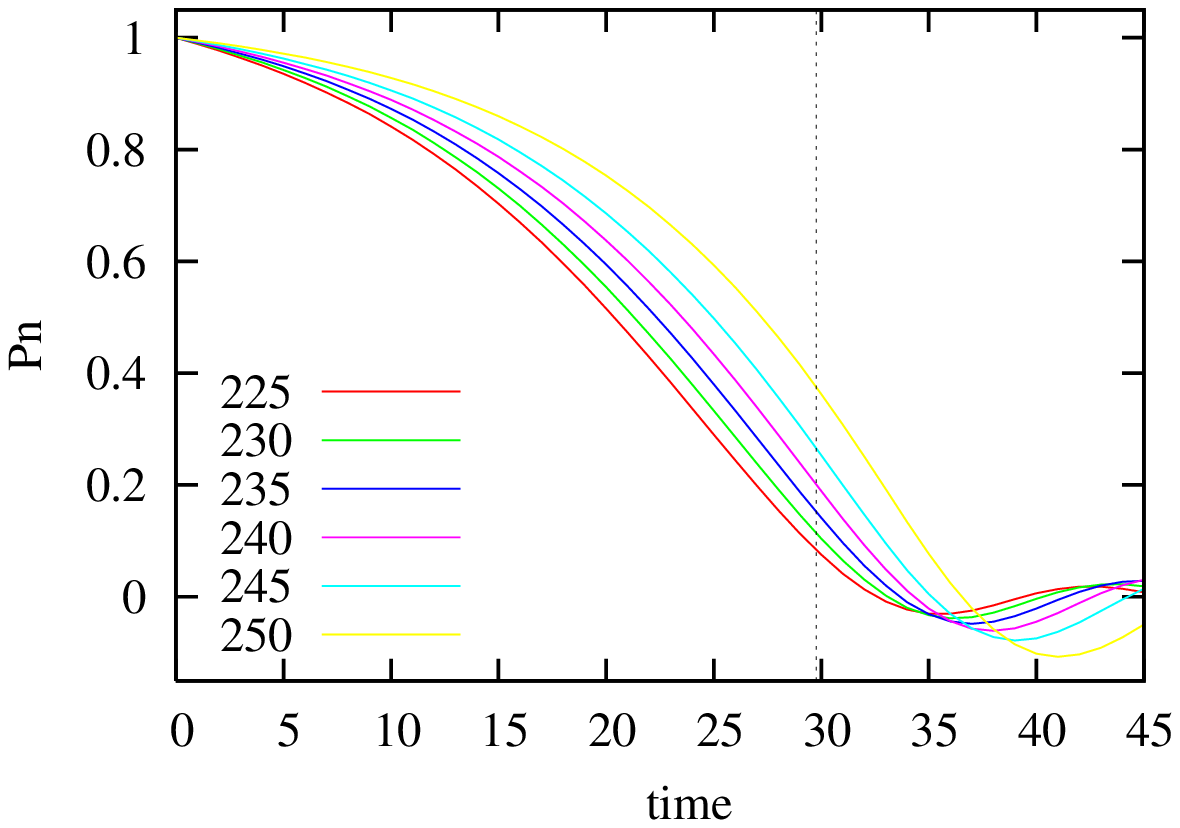, angle = -90, width = 8.1cm}%
\end{center}
\caption{The evolution of the fields. In the left panel, masses are distributed
according to Eq.~(\ref{logdist}) where the most massive mass $m_1 = m_\mathrm{Pl}$,
while in the right panel Eq.~(\ref{MPdist}) is applied with the average mass
$\bar{m} = m_\mathrm{Pl}$ and the parameter $\beta = 1/2$. As in
Fig.~\ref{singleplot}, the vertical dotted line denotes the end of inflation.}%
\label{fieldplot}
\end{figure}

We have performed numerical calculations under this mass distribution with the
lightest mass being 1/100 of the heaviest mass, i.e. $m_N = 10^{-2} m_1$. We have
tried $N = 250$, 500 and 1000, with $m_1 = m_\mathrm{Pl}$, $10^{-2} m_\mathrm{Pl}$,
$10^{-4} m_\mathrm{Pl}$ and $10^{-6} m_\mathrm{Pl}$ for each case. The evolution of
several light fields is shown in the left panel of Fig.~\ref{fieldplot}. Here, we
show only the case where we have set $m_1 = m_\mathrm{Pl}$ and $N = 250$, since for
other cases the fields follow exactly the same evolutions except for time scale:
e.g. it takes $10^2$ times longer when $m_1$ becomes $10^{-2}$ times lighter. The
dotted line indicates the end of inflation. As can be seen, heavy fields drop out of
the inflationary regime at early times and only a limited number of light fields
keep non-zero amplitudes at the end of inflation. To estimate the number of the
fields which contribute at the end of inflation, we can combine
Eqs.~(\ref{heaviest}) and (\ref{logdist}) to obtain
\begin{equation}
n \sim 1 + \frac{(N - 1)\log 2}{4 \log 10} \, ,
\end{equation}
which gives $n = 19$, 38 and 76 for $N = 250$, 500 and 1000, respectively. Also note
that because of the relatively large $H$, many fields undergo dissipation
\cite{dkmw} at the early stage of inflation, and do not oscillate near the minima.

\begin{figure}[h]
\psfrag{P}{$\mathcal{P}$}%
\psfrag{N}{$\mathcal{N}$}%
\psfrag{m1isPl-0}{$m_1 = m_\mathrm{Pl}$}%
\psfrag{m1isPl-2}{$m_1 = m_\mathrm{Pl}^{-2}$}%
\psfrag{m1isPl-4}{$m_1 = m_\mathrm{Pl}^{-4}$}%
\psfrag{m1isPl-6}{$m_1 = m_\mathrm{Pl}^{-6}$}%
\psfrag{mbisPl-0}{$\bar{m} = m_\mathrm{Pl}$}%
\psfrag{mbisPl-2}{$\bar{m} = m_\mathrm{Pl}^{-2}$}%
\psfrag{mbisPl-4}{$\bar{m} = m_\mathrm{Pl}^{-4}$}%
\psfrag{mbisPl-6}{$\bar{m} = m_\mathrm{Pl}^{-6}$}%
\begin{center}
\epsfig{file=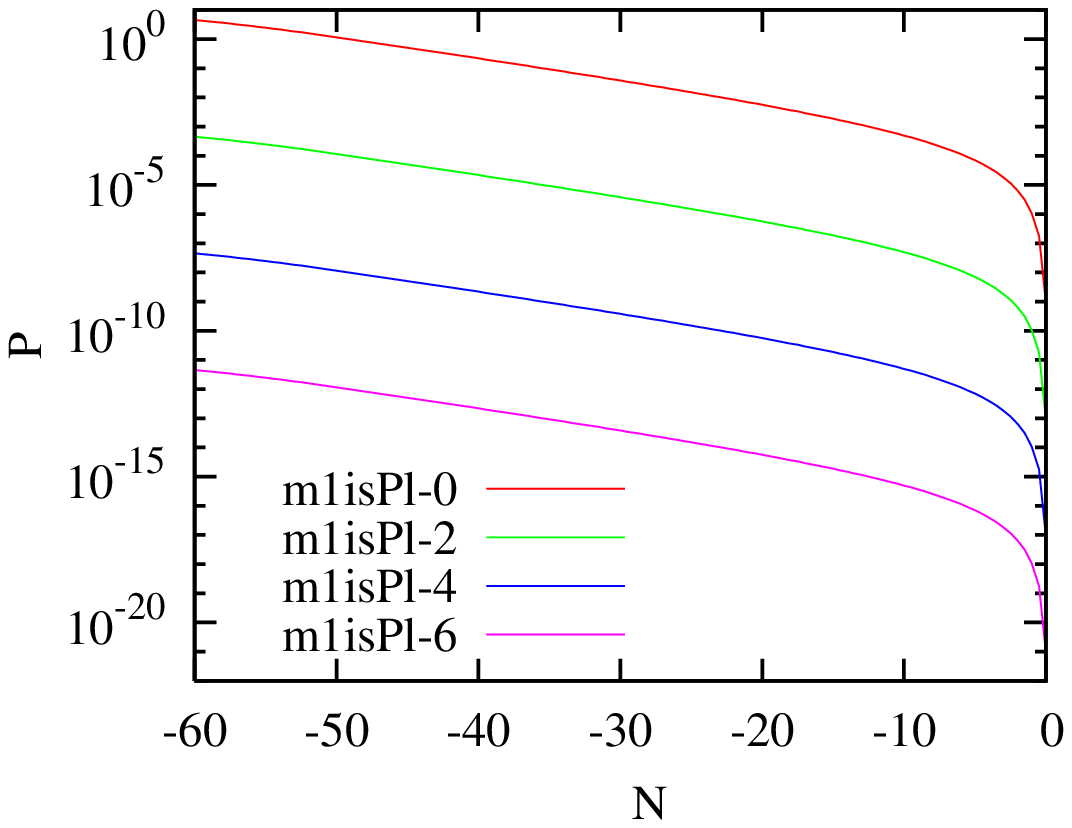, angle = -90, width = 8.1cm}%
\epsfig{file=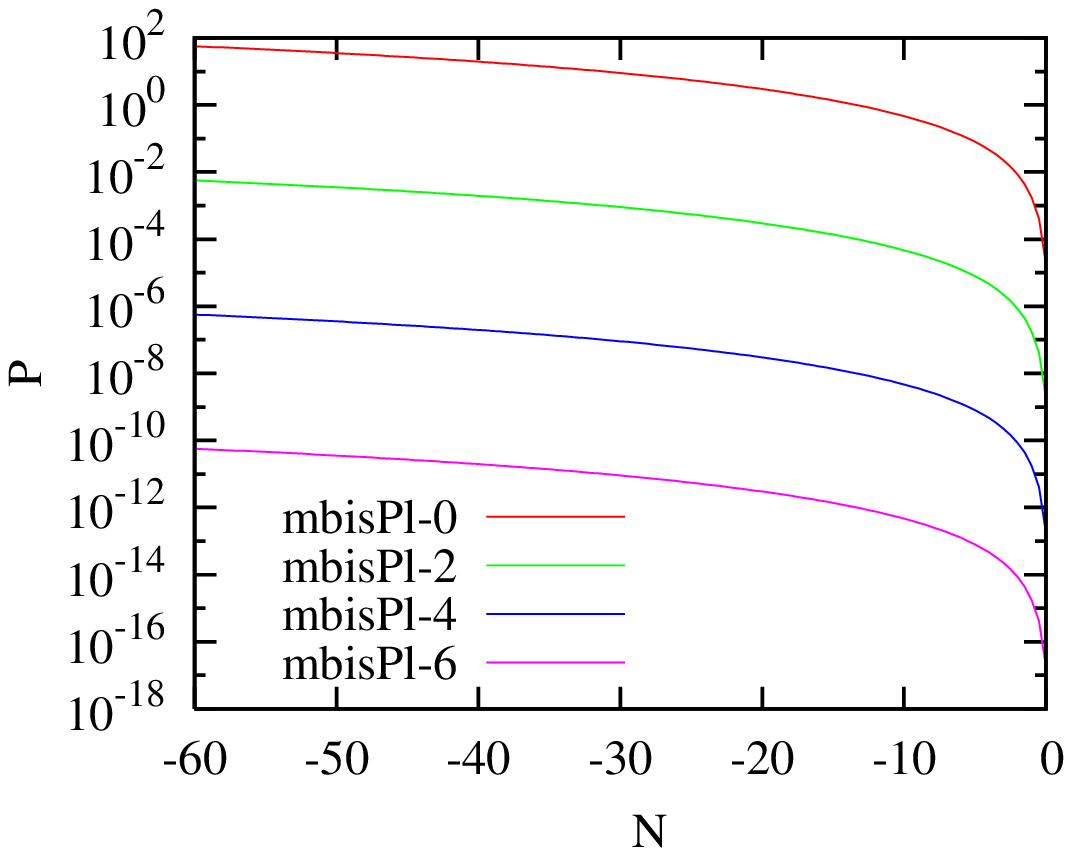, angle = -90, width = 8.1cm}%
\end{center}
\caption{Plots of the power spectra $\mathcal{P}$ calculated by Eq.~(\ref{spectrum})
corresponding to the calculations in Fig.~\ref{fieldplot}. We follow the mass
distribution given by Eqs.~(\ref{logdist}) and (\ref{MPdist}) in the left and right
panel, respectively.}%
\label{spectrumplot}
\end{figure}

In the left panel of Fig.~\ref{spectrumplot} we show the power spectrum
$\mathcal{P}$ given by Eq.~(\ref{spectrum}) for this case. Now, we obtain different
spectra for different cases: the mass gives overall amplitude, and the number of
fields controls the shape of the spectrum, as we will see soon. As can be read from
Eq.~(\ref{spectrum}), the amplitude of the spectrum is proportional to the sum of
the mass squared of every field. We can also understand the reason why the spectrum
becomes flatter with larger number of fields as follows: since here we take
$\phi_i^{(i)} = m_\mathrm{Pl}$ for all $i$, massive fields dominate the total energy
density at early times. Therefore as these fields are completely dropping out of the
inflationary regime, the Hubble parameter $H$ is decreasing steeply, making the
amplitude $\mathcal{P}$ decline quickly. This ``drop-out'' effect of heavy fields
weakens when there are many other light fields so the contributions by these fields
keep the energy density and hence $H$ smoothly varying within the slow-roll regime.
This is independent of the overall mass scale but dependent only on the number of
fields and the spacing between masses. Thus, all in all, we can observe the same
shape of the spectrum with the same number of fields.

\begin{figure}[h]
\psfrag{n}{$n_\mathrm{s}$}%
\psfrag{N}{$\mathcal{N}$}%
\psfrag{Nis-250}{$N = 250$}%
\psfrag{Nis-500}{$N = 500$}%
\psfrag{Nis-000}{$N = 1000$}%
\begin{center}
\epsfig{file=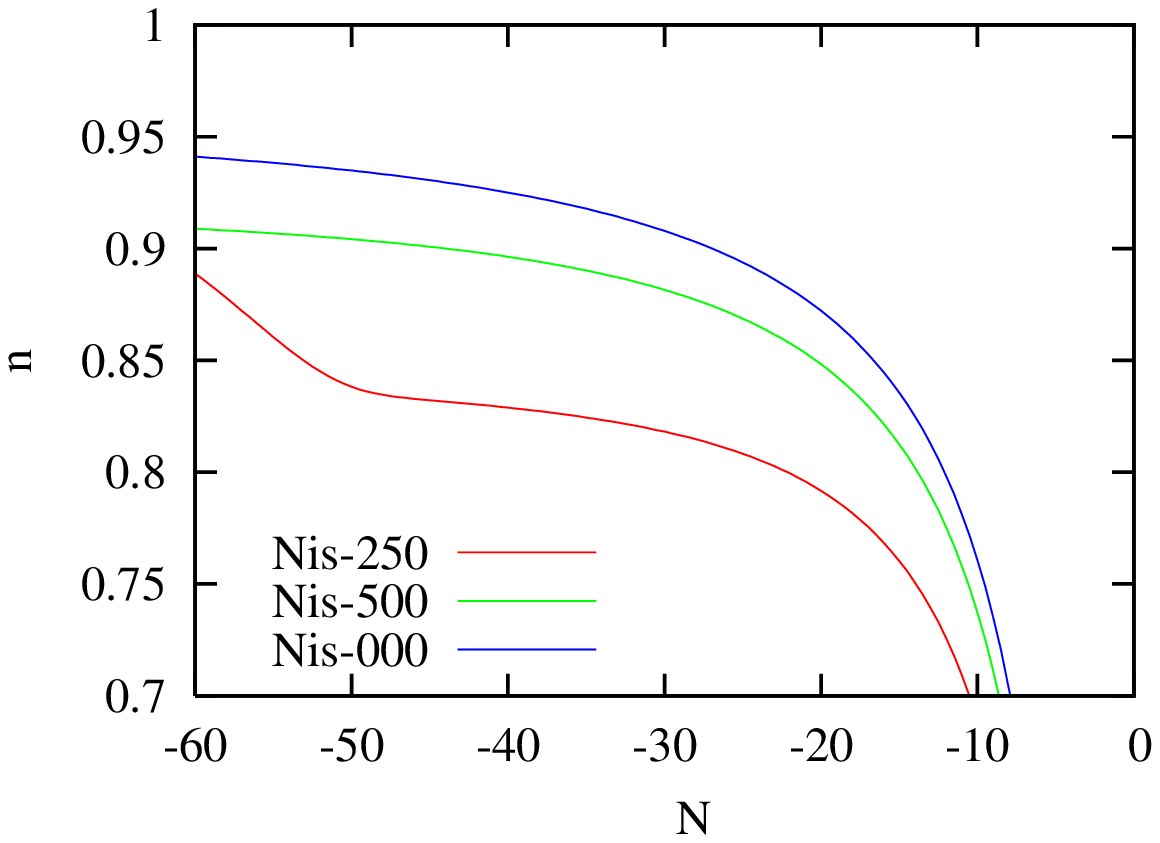, angle = -90, width = 8.1cm}%
\epsfig{file=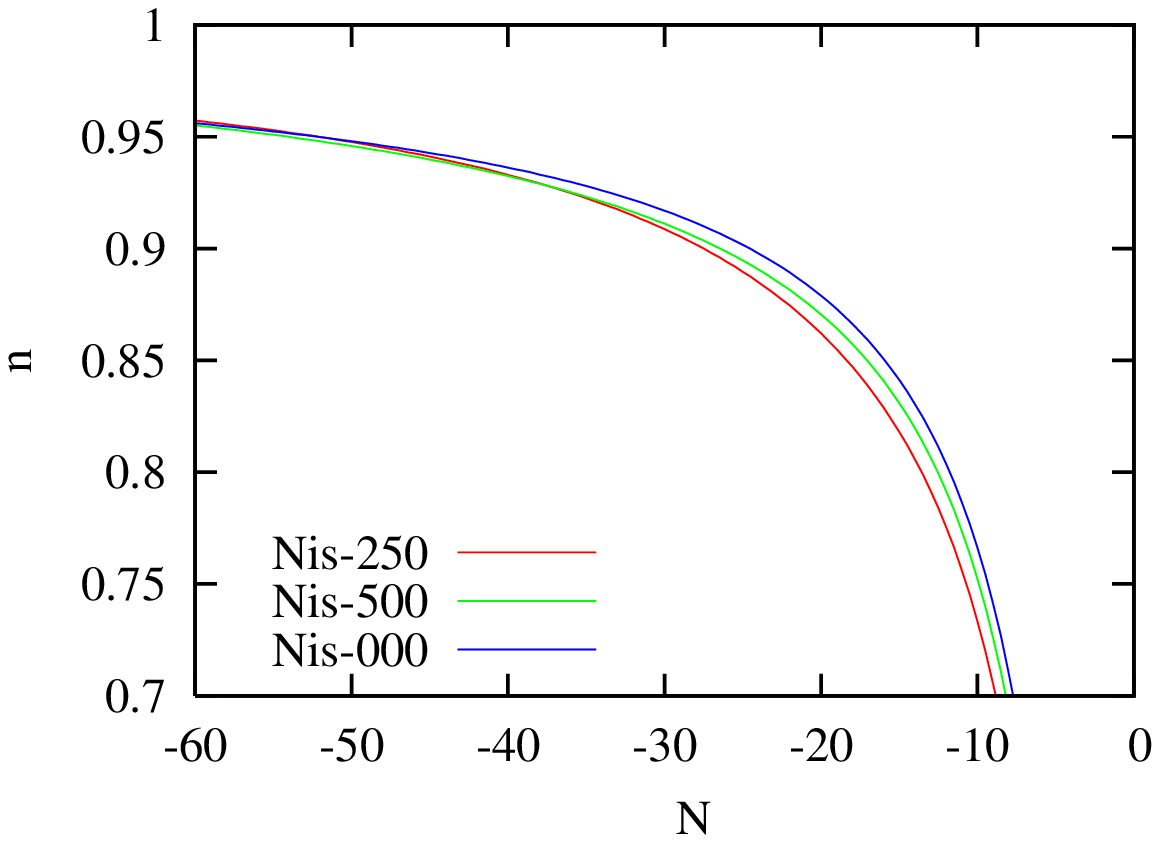, angle = -90, width = 8.1cm}
\end{center}
\caption{The spectral indices $n_\mathrm{s}$ of the spectra shown in
Fig.~\ref{spectrumplot}. The shape of the spectrum is identical irrespective of the
overall mass scale. In the left panel, the mass distribution follows
Eq.~(\ref{logdist}), with the lightest mass being 1/100 of the heaviest one: for the
case of $N = 250$ where we can obtain $\mathcal{N} \sim 60$, $n_\mathrm{s}$ is
decreasing quickly at the early stages due to the ``drop-out'' effect of the massive
fields. For the Mar{\v c}enko-Pastur law given by Eq.~(\ref{MPdist}) where the
masses are much more densely spaced, shown in the right panel, this effect is very
weak and $n_\mathrm{s}$ looks more or less the same at 60 $e$-folds before the end
of inflation. Still we can see that the index for $N = 250$ case is redder than
$n_\mathrm{s}$ for $N = 500$ and $N = 1000$ cases.}%
\label{indexplot}
\end{figure}

Now we turn to the spectral index $n_\mathrm{s}$. As we have discussed above, here
$n_\mathrm{s}$ is independent of the overall mass scale, so in the left panel of
Fig.~\ref{indexplot} we show different indices on different $N$ only. As anticipated
from the discussion above, we obtain larger tilt with smaller number of fields. One
interesting thing is the initial steep decrease of the index for $N = 250$. This is
due to the drop-out effect we discussed above: massive fields which occupy a large
fraction of the energy density drop out of the inflationary regime, so $H$, and
consequently $\mathcal{P}$ is decreasing quickly. However, soon the contributions by
lighter fields under the slow-roll evolution become dominant and there is no more
rapid change in $H$, so $\mathcal{P}$ becomes flatter afterwards. Therefore, because
of this drop-out effect, typically the spectrum is redder than the spectrum of the
single field inflation where such an effect cannot happen.

\subsection{Mar{\v c}enko-Pastur law}
\label{subsecMPdist}

The mass of string axion depends on the details of the compactification which
involves many factors not yet (or formidable to be) computed. Nevertheless, in spite
of the unclear microscopic physics, the mass distribution is known to follow the
Mar{\v c}enko-Pastur law \cite{mpdist}: the probability density function for $m^2$
is given by
\begin{equation}\label{MPdist}
p(m^2) = \frac{\sqrt{ \left[ \bar{m}^2 \left( 1 + \sqrt{\beta} \right)^2 - m^2
\right] \left[ m^2 - \bar{m}^2 \left( 1 - \sqrt{\beta} \right)^2 \right]
}}{2\pi\beta \bar{m}^2 m^2} \, ,
\end{equation}
where $\bar{m}$ is the average value of the mass, i.e. $\langle m^2 \rangle =
\bar{m}^2$, and $\beta$ is a model dependent parameter whose favourable value is
expected to be 1/2. Note that for this case masses are more densely packed than the
exponential distribution we have used in the previous section: we have
\begin{align}
m_1 & = 1.707110 \bar{m} \, ,
\nonumber \\
m_N & = 0.292893 \bar{m}
\end{align}
for $\beta = 1/2$.

The evolution the of fields according to the Mar{\v c}enko-Pastur law given by
Eq.~(\ref{MPdist}) is shown in the right panel of Fig.~\ref{fieldplot}. Since masses
are densely packed, the fields are evolving faster than those under the distribution
Eq.~(\ref{logdist}) we have used in the previous section, where the masses spread
over two orders of magnitude. The qualitative evolution of the fields is, however,
not too different: heavy fields exit the inflationary regime at early times, and
light fields drive inflation at late stages.

The relative dense packing of the masses is easily read from the power spectrum
$\mathcal{P}$ and the spectral index $n_\mathrm{s}$, which are shown in the right
panels of Figs.~\ref{spectrumplot} and \ref{indexplot}, respectively. In
Fig.~\ref{spectrumplot}, we can see that the amplitudes of the spectra are
relatively large and flat compared with the corresponding curves in the left panel
where the exponential distribution of the previous section, Eq.~(\ref{logdist}), is
applied. Larger amplitude is easily anticipated from the fact that masses are
densely spaced: the most massive field is 100 times heavier than the lightest one in
the left panel, while now it is less than 6 times. Hence the energy density is large
with the same initial conditions thus correspondingly $\mathcal{P}$ is also large.
Flatter spectrum could be understood in the same way: since the masses are very
close to each other, the drop-out effect is weak and hence the spectra with
different numbers of fields look very similar. This is clearly seen by the spectral
index shown in Fig.~\ref{indexplot}. Unlike the case of the exponential distribution
shown in the left panel, $n_\mathrm{s}$'s look very similar now. Still,
$n_\mathrm{s}$ for the case of 250 fields is slightly redder than the other cases
where more fields support inflation so the effect of the slow-roll is more powerful.
We have summarised the results of the current and the previous sections in
Table~\ref{bigtable}.

\begin{sidewaystable}[h]
\begin{tabular}{r||*{9}{l|}l}
 mass & $N$ & m & \multicolumn{2}{|l}{$\mathcal{N}$} &
 \multicolumn{2}{|l}{$\phi_N^{\mathrm{end}}$} &
 \multicolumn{2}{|l|}{$H_\mathrm{end}$} & $\mathcal{P}$ & $n_\mathrm{s}$
 \\
 \cline{4-9}
 distribution & & & analytic & numerical & analytic & numerical & analytic &
 numerical & &
 \\
 \hline \hline
 & 250 & $m_\mathrm{Pl}$ & 62.50 & 64.52 & 0.31427 & 0.353184 & $0.471405 \times
 10^{-2}$ & $0.604933 \times 10^{-2}$ & $4.82272 \times 10^{0}$ & 0.891368
 \\
 & & $m_\mathrm{Pl}^{-2}$ & & & & & $0.471405 \times 10^{-4}$ & $0.604933 \times
 10^{-4}$ & $4.82272 \times 10^{-4}$&
 \\
 & & $m_\mathrm{Pl}^{-4}$ & & & & & $0.471405 \times 10^{-6}$ & $0.604933 \times
 10^{-6}$ & $4.82272 \times 10^{-8}$&
 \\
 & & $m_\mathrm{Pl}^{-6}$ & & & & & $0.471405 \times 10^{-8}$ & $0.604933 \times
 10^{-8}$ & $4.82272 \times 10^{-12}$&
 \\
 \cline{2-11}
 & 500 & $m_\mathrm{Pl}$ & 125.00 & 127.13 & 0.219199 & 0.271686 &
 $0.471405 \times 10^{-2}$ & $0.588379 \times  10^{-2}$ & $1.80712 \times 10^{-1}$ &
 0.909015
 \\
 exponential & & $m_\mathrm{Pl}^{-2}$ & & & & & $0.471405 \times 10^{-4}$ & $0.588379
 \times 10^{-4}$ & $1.80712 \times 10^{-5}$&
 \\
 distribution & & $m_\mathrm{Pl}^{-4}$ & & & & & $0.471405 \times 10^{-6}$ &
 $0.588379 \times 10^{-6}$ & $1.80712 \times 10^{-9}$&
 \\
 & & $m_\mathrm{Pl}^{-6}$ & & & & & $0.471405 \times 10^{-8}$ & $0.588379 \times
 10^{-8}$ & $1.80712 \times 10^{-13}$&
 \\
 \cline{2-11}
 & 1000 & $m_\mathrm{Pl}$ & 250.00 & 252.24 & 0.153960 & 0.206531 & $0.471405 \times
 10^{-2}$ & $0.576077 \times 10^{-2}$ & $3.71709 \times 10^{-2}$ & 0.941230
 \\
 & & $m_\mathrm{Pl}^{-2}$ & & & & & $0.471405 \times 10^{-4}$ & $0.576077 \times
 10^{-4}$ & $3.71709 \times 10^{-6}$&
 \\
 & & $m_\mathrm{Pl}^{-4}$ & & & & & $0.471405 \times 10^{-6}$ & $0.576077 \times
 10^{-6}$ & $3.71709 \times 10^{-10}$&
 \\
 & & $m_\mathrm{Pl}^{-6}$ & & & & & $0.471405 \times 10^{-8}$ & $0.576077 \times
 10^{-8}$ & $3.71709 \times 10^{-14}$&
 \\
 \hline
 & 250 & $m_\mathrm{Pl}$ & 62.50 & 63.69 & 0.305888 & 0.375010 & $0.142110 \times
 10^{0}$ & $0.196380 \times 10^{0}$ & $5.74883 \times 10^{1}$ & 0.957510
 \\
 & & $m_\mathrm{Pl}^{-2}$ & & & & & $0.142110 \times 10^{-2}$ & $0.196380 \times
 10^{-2}$ & $5.74883 \times 10^{-3}$&
 \\
 & & $m_\mathrm{Pl}^{-4}$ & & & & & $0.142110 \times 10^{-4}$ & $0.196380 \times
 10^{-4}$ & $5.74883 \times 10^{-7}$&
 \\
 & & $m_\mathrm{Pl}^{-6}$ & & & & & $0.142110 \times 10^{-6}$ & $0.196380 \times
 10^{-6}$ & $5.74883 \times 10^{-11}$&
 \\
 \cline{2-11}
 Mar{\v c}enko- & 500 & $m_\mathrm{Pl}$ & 125.00 & 126.32 & 0.222222 & 0.301697 &
 $0.140584 \times 10^{0}$ & $0.188119 \times 10^{0}$ & $3.82160 \times 10^{1}$ &
 0.955256
 \\
 Pastur & & $m_\mathrm{Pl}^{-2}$ & & & & & $0.140584 \times 10^{-2}$ & $0.188119
 \times 10^{-2}$ & $3.82160 \times 10^{-3}$ &
 \\
 distribution & & $m_\mathrm{Pl}^{-4}$ & & & & & $0.140584 \times 10^{-4}$ &
 $0.188119 \times 10^{-4}$ & $3.82160 \times 10^{-7}$ &
 \\
 & & $m_\mathrm{Pl}^{-6}$ & & & & & $0.140584 \times 10^{-6}$ & $0.188119 \times
 10^{-6}$ & $3.82160 \times 10^{-11}$&
 \\
 \cline{2-11}
 & 1000 & $m_\mathrm{Pl}$ & 250.00 & 251.45 & 0.159364 & 0.239128 & $0.139643 \times
 10^{0}$ & $0.182052 \times 10^{0}$ & $2.45069 \times 10^{1}$ & 0.956110
 \\
 & & $m_\mathrm{Pl}^{-2}$ & & & & & $0.139643 \times 10^{-2}$ & $0.182052 \times
 10^{-2}$ & $2.45069 \times 10^{-3}$ &
 \\
 & & $m_\mathrm{Pl}^{-4}$ & & & & & $0.139643 \times 10^{-4}$ & $0.182052 \times
 10^{-4}$ & $2.45069 \times 10^{-7}$ &
 \\
 & & $m_\mathrm{Pl}^{-6}$ & & & & & $0.139643 \times 10^{-6}$ & $0.182052 \times
 10^{-6}$ & $2.45069 \times 10^{-11}$ &
\end{tabular}
\caption{Summary of the numerical calculations we have performed in
Sections~\ref{subseclogdist} and \ref{subsecMPdist}. For analytic estimates, we have
used the results derived in Section~\ref{secNflation}. Note that the mass given here
denotes the heaviest mass, $m_1$, and the average mass $\bar{m}$ for the exponential
distribution and the Mar{\v c}enko-Pastur law, respectively. Also,
$\phi_\mathrm{end}$ and $H_\mathrm{end}$ are given in term of $m_\mathrm{Pl}$, and
$\mathcal{P}$ and $n_\mathrm{s}$ are evaluated at 60 $e$-folds before the end of
inflation.}%
\label{bigtable}
\end{sidewaystable}

\subsection{Comparison with single field case}

As we have seen in the previous sections, the number of the axion fields and the
corresponding mass distributions have important effects on the power spectrum and
the spectral index. And this affects the physical quantities at the end of the
inflationary epoch we have calculated in Section~\ref{secNflation}. To get a clearer
idea, in this section we compare two mass distributions we have discussed in the
previous sections for multi-field inflation with the case of single field inflation.

\begin{figure}[h]
\psfrag{P}{$\mathcal{P}$}%
\psfrag{ns}{$n_\mathrm{s}$}%
\psfrag{N}{$\mathcal{N}$}%
\begin{center}
\epsfig{file=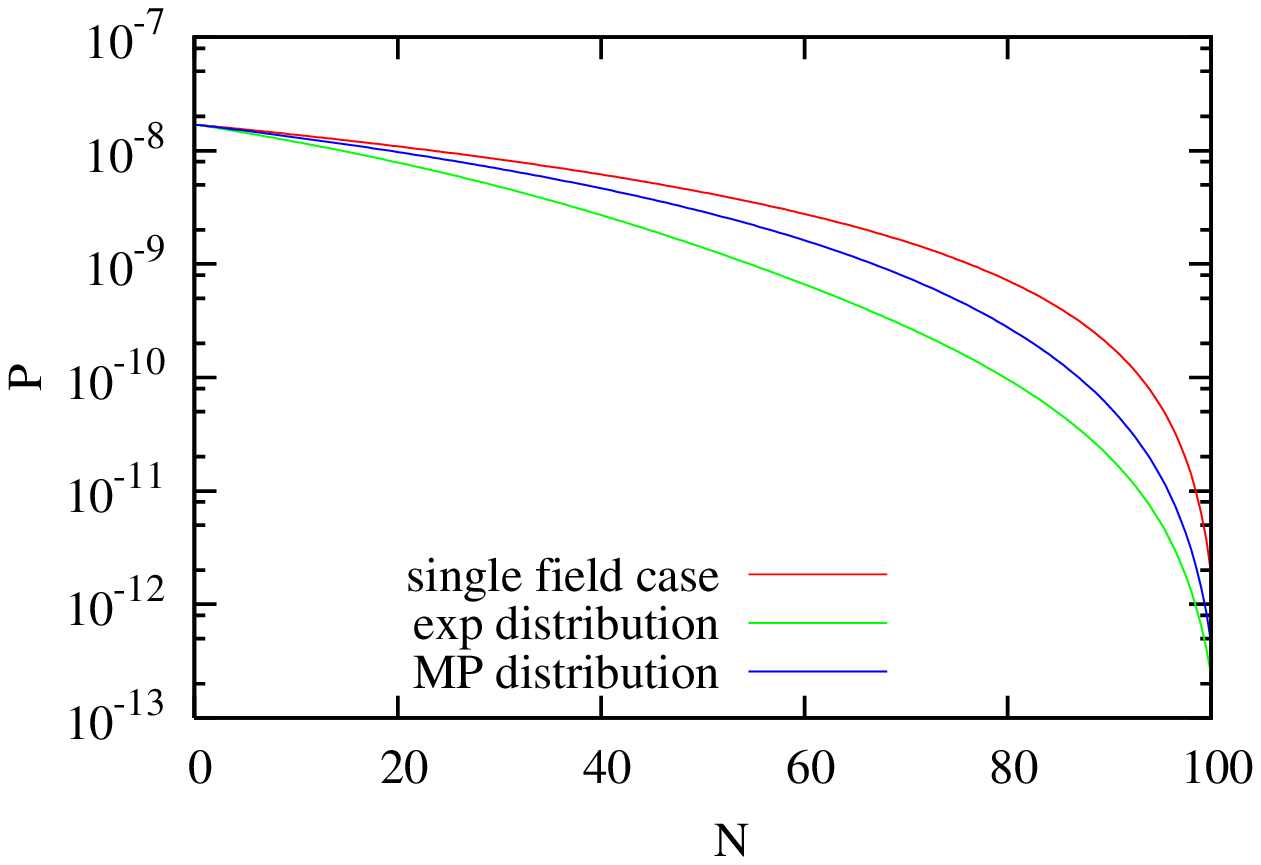, angle = -90, width = 8.1cm}%
\epsfig{file=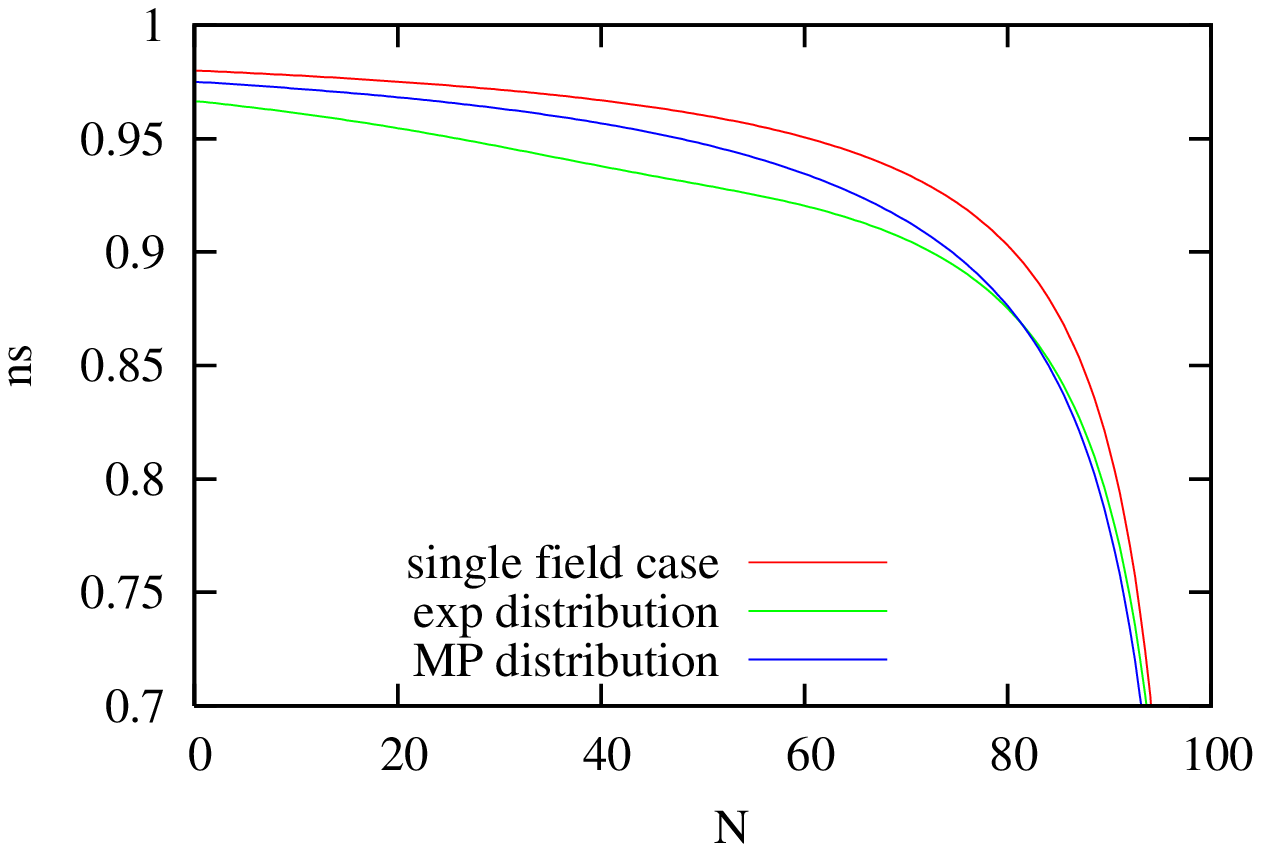, angle = -90, width = 8.1cm}%
\end{center}
\caption{Plots of the power spectra $\mathcal{P}$ (left panel) and the spectral
indices $n_\mathrm{s}$ (right panel). Here we set $\mathcal{N} = 0$ to be the
beginning of the inflationary phase. At $\mathcal{N} = 0$, the amplitudes of the
power spectra for all the three cases are the same, but the spectral indices are
not: this is because for the exponential distribution the masses are sparsely placed
in heavy region, so the drop-out effect is larger at early times. Note that since
the masses are densely packed in light region for the exponential distribution, more
fields support inflation at late stages than for the Mar{\v c}enko-Pastur
distribution and hence $n_\mathrm{s}$ becomes larger before the end of inflation.}%
\label{compareplot}
\end{figure}

We have performed numerical calculations on the basis of the single field inflation
of Section~\ref{secsingle}, where we set $\phi_i = 20 m_\mathrm{Pl}$ and $m =
10^{-5} m_\mathrm{Pl}$. We have tried total 400 fields for multi-field cases, and
for comparison we have set the average mass the same: for the exponential
distribution, Eq.~(\ref{logdist}),
\begin{equation}
m_1 = 10 m_{400} = 2.15313 \times 10^{-5} m_\mathrm{Pl} \, ,
\end{equation}
i.e. the lightest mass is 1/10 of the heaviest one. We simply set $\bar{m} = 10^{-5}
m_\mathrm{Pl}$ for the Mar{\v c}enko-Pastur law, Eq.~(\ref{MPdist}), then we have
\begin{align}
m_1 & = 1.69594 \times 10^{-5} m_\mathrm{Pl} \, ,
\nonumber \\
m_{400} & = 0.299126 \times 10^{-5} m_\mathrm{Pl} \, .
\end{align}
We have set all the initial amplitudes the same,
\begin{align}
\phi_i^{(0)} & = 1.00000 m_\mathrm{Pl} \, ,
\nonumber \\
\phi_i^{(0)} & = 0.999968 m_\mathrm{Pl}
\end{align}
for the exponential distribution and the Mar{\v c}enko-Pastur distribution
respectively, making the initial Hubble parameters $H_\mathrm{ini}$'s for all the
three cases identical. In Fig.~\ref{compareplot} and Table~\ref{comparetable}, we
show the results of the calculations.

In the left panel of Fig.~\ref{compareplot}, we plot the power spectra
$\mathcal{P}$. Since $H_\mathrm{ini}$'s are the same, the amplitude of
$\mathcal{P}$'s are identical initially. However, as inflation proceeds they drop at
different rates: $\mathcal{P}$ for the single field case is most slowly evolving,
while for multi-field cases we obtain steeper spectra. Here, how quickly the
spectrum is decreasing depends on the density of the masses: if the mass density is
high, i.e. masses of the fields are densely packed, the drop-out effect is
alleviated and the spectrum is slowly changing compared with the case of sparse mass
density. This is clearly seen in the right panel of Fig.~\ref{compareplot} which
shows the spectral indices $n_\mathrm{s}$. Note that at later stages, around
$\mathcal{N} \sim 80$, $n_\mathrm{s}$ under the Mar{\v c}enko-Pastur law becomes
steeper than the one under the exponential distribution: following
Eq.~(\ref{logdist}), the masses of the fields are densely placed in the lighter
region. Thus at later stages where light fields support inflation the energy density
under the exponential distribution is changing more mildly, making the spectrum
bluer.

\section{Conclusions}
\label{secconclusion}

In this paper we have examined the dynamics of multi-field inflation at its final
stage and the density perturbations. Within the slow-roll regime, the fields follow
the evolution equations Eq.~(\ref{fieldrelation}) according to their masses and we
can write all the fields and their derivatives in terms of the lightest field as
Eq.~(\ref{phiNrelation}). Then, the Hubble parameter at the end of inflation,
$H_\mathrm{end}$, becomes independent of many factors, e.g. initial values and the
number of the fields and the detail of the mass distribution, and is approximated to
the lightest mass $m_N$ only, given by Eq.~(\ref{Hendsimple}). Also the amplitude of
the lightest field at the end of inflation, $\phi_N^\mathrm{end}$, depends on the
number of fields which have not yet relaxed to the minima of the potentials as
Eq.~(\ref{phiNend}). The power spectrum $\mathcal{P}$ and the spectral index
$n_\mathrm{s}$ given by Eqs.~(\ref{spectrum}) and (\ref{index}) respectively,
however, cannot be calculated in the same way because many fields other than the
lightest one contribute at earlier stages of inflation.

The detail of the mass distribution greatly affects $\mathcal{P}$ and
$n_\mathrm{s}$: we have investigated the exponential distribution,
Eq.~(\ref{logdist}), and the Mar{\v c}enko-Pastur law, Eq.~(\ref{MPdist}). It seems
that three factors of the mass distribution, namely, the number of the fields $N$,
the overall mass scale, and the density of masses in the range over which they are
spread. The amplitude of the power spectrum $\mathcal{P}$ depends on the overall
mass scale, and the spectral index $n_\mathrm{s}$ is independent of the mass scale
but is affected by the two other factors. The ``drop-out'' effect is crucial here:
the most massive axions fall to their minima at the early stages of inflation and
their energies are dissipated by the large Hubble parameter, completely dropping out
of the inflationary regime and contribute no more. This leads to the relatively
steep decrease in $H$ and hence makes $n_\mathrm{s}$ redder than the single field
case. If the masses are densely packed and the number of the fields are large, the
drop-out effect is alleviated because soon the slowly rolling fields dominate the
energy density.

Note that, however, detail of the initial conditions and the masses of the axion
fields depend on the structure of the microphysics of string compactification.
Although it is a formidable task to calculate any specific and concrete realisation
from the first principle, this would be an important subject to implement a
consistent and successful scenario of inflation in the context of string theory.

\subsection*{Acknowledgements}

I thank Kiwoon Choi, Hongsu Kim and Soo A Kim for useful conversations. Especially I
am indebted to Ki-Young Choi and Andrew Liddle for many important correspondences
and suggestions. It is also a great pleasure to thank Seung-Hoon Cha, Sungwook Hong,
Donghui Jeong and Kwangil Seon for helpful comments on numerical computations.

\end{document}